\documentclass[a4paper,11pt]{article}
\pdfoutput=1 % if your are submitting a pdflatex (i.e. if you have
\usepackage{jinstpub} % for details on the use of the package, please
\usepackage{graphicx}
\usepackage{subfigure}
\usepackage{float}
\usepackage{lineno}

\title{\boldmath Capability of detecting low energy events in JUNO Central Detector}

\author[a]{X. Fang}
\author[b]{Y. Zhang}
\author[d]{G.H. Gong}
\author[c,1,2]{G.F. Cao\note{Corresponding author.}\note{Also at University of Chinese Academy of Sciences, Beijing 100049, China.}}
\author[c]{T. Lin}
\author[a]{C.W. Yang}
\author[c]{W.D. Li}
\affiliation[a]{School of Physics, Sichuan University, Chengdu 610065, China}
\affiliation[b]{University of Chinese Academy of Sciences, Beijing 100049, China}
\affiliation[c]{Institute of High Energy Physics, Chinese Academy of Sciences, Beijing 100049, China}
\affiliation[d]{Tsinghua Uviversity, Beijing 100084, China}

\abstract{The Jiangmen Underground Neutrino Observatory (JUNO) is an experimental project designed to determine the neutrino mass ordering and probe the fundamental properties of the neutrino oscillations. The JUNO central detector is a spherical liquid scintillator detector with a diameter of 35.4 m and equipped with approximately 18,000 20-inch PMTs. A trigger threshold of 0.5 MeV can be easily achieved by using a common multiplicity trigger and can meet the requirements for measuring neutrino mass ordering. However, it is essential to further reduce the trigger threshold for detecting solar neutrinos and supernova neutrinos. A sophisticated trigger scheme is proposed to achieve a low energy threshold by reducing the level of low energy radioactivity and dark noise coincidence. With the new trigger scheme, the events rate of the central detector from different types of sources have been carefully studied by using a detailed detector simulation. It shows that the trigger threshold can be reduced to 0.2 MeV, or even 0.1 MeV, if the concentration of $^{14}$C in liquid scintillator can be well controlled.}


\begin{document}
%\linenumbers
\maketitle
%\flushbottom
	
\section{Introduction}
The Jiangmen Underground Neutrino Observatory (JUNO) is a multipurpose neutrino experiment which is mainly designed to determine neutrino mass ordering and precisely measure oscillation parameters \cite{lab1}. Other physics topics include observing supernova neutrinos, studying atmospheric neutrinos, solar neutrinos and geo-neutrinos, and performing exotic searches, etc. It is located in Jiangmen in southern China, about 53 km away from Yangjiang and Taishan nuclear power plants. The experiment is now under construction and data collection is expected to begin in 2021.
		
For the JUNO experiment, electrical signals from the photo-sensors are firstly acquired, amplified and digitized by the front-end electronics and then part of them containing interesting events will be selected by the trigger system in an efficient and controlled way, and finally data acquisition system moves the data for those events to persistent storage for later offline data processing and analysis. In this process, the trigger system with a low energy threshold is desired and essential to precisely measure $^{7}$Be and pp solar neutrinos and supernova neutrinos with JUNO central detector. Measurements of $^{7}$Be and pp solar neutrinos can significantly improve our knowledge on solar physics. In the central detector, the flux and energy spectra of $^{7}$Be and pp solar neutrinos can be measured via the elastic neutrino electron scattering, where the energy of recoil electrons are continuous and the end points of their energy spectra are $\sim$0.75~MeV and $\sim$0.3~MeV for $^{7}$Be and pp neutrinos, respectively. Therefore, a trigger threshold of 0.2~MeV or even lower is preferable, in order to accumulate sufficient statistics. This situation is also true for detection channels of elastic neutrino scattering off electrons and protons for supernova neutrinos. Particularly for recoil protons, their energies are quenched in liquid scintillator, which yields a much lower visible energy. In this paper, we propose a new trigger strategy to detect low energy events, and this trigger strategy is studied by using the full detector simulation software. Based on the proposed trigger scheme, the events rate of JUNO central detector is estimated, which is limiting the trigger threshold and being an important input to design electronics, trigger and data acquisition systems (DAQ). It is also one of the major parameters determining the scale of computing facility for storage and computing.
                
This paper is organized as follows: at first the layout of JUNO detector is introduced, followed by a description of the trigger schemes. Then the performance of the trigger schemes is presented based on the full detector simulation. Finally, the events rate from different types of sources, such as physical events, cosmic muons, radioactivity background, etc, are given for the most favorable configuration.

\section{JUNO detector}
The JUNO detector \cite{lab2} is composed of  Central Detector (CD), Water Cherenkov Detector (WCD) and Top Tracker (TT) as shown in Fig. 1. The CD is made of 20 kt liquid scintillator contained in a 35.4-meter diameter acrylic sphere with the designed energy resolution of 3\% at 1 MeV, corresponding to the energy scale of $\sim$1200 pe/MeV. The acrylic sphere is supported by a stainless steel truss, on which $\sim$18,000 20-inch PMTs are mounted to detect scintillation lights emitting from the LS. The coverage of photo-cathode is 75\% achieved by carefully investigating the arrangement of PMTs. 5,000 PMTs are dynode PMTs produced by Hamamatsu with better transit time spread and others are new type micro channel plate PMTs (MCP-PMT) manufactured by Northern Night Vision Technology (NNVT) in China. The acrylic sphere will be immersed in a water pool used to shield the background of natural radioactivity from surrounding rocks and environment. The water pool also works as a water cherenkov detector which can identify cosmic muons by detecting their cherenkov lights. The water buffer between the PMTs and the acrylic sphere can also shield the radioactivity background from the PMTs and stainless steel truss. The thickness of the buffer is about 1.4 meters. On the top of the CD, a muon tracker is installed to detect muons with better resolution.
                
\begin{figure}[H]
\centering
\includegraphics[width=7cm]{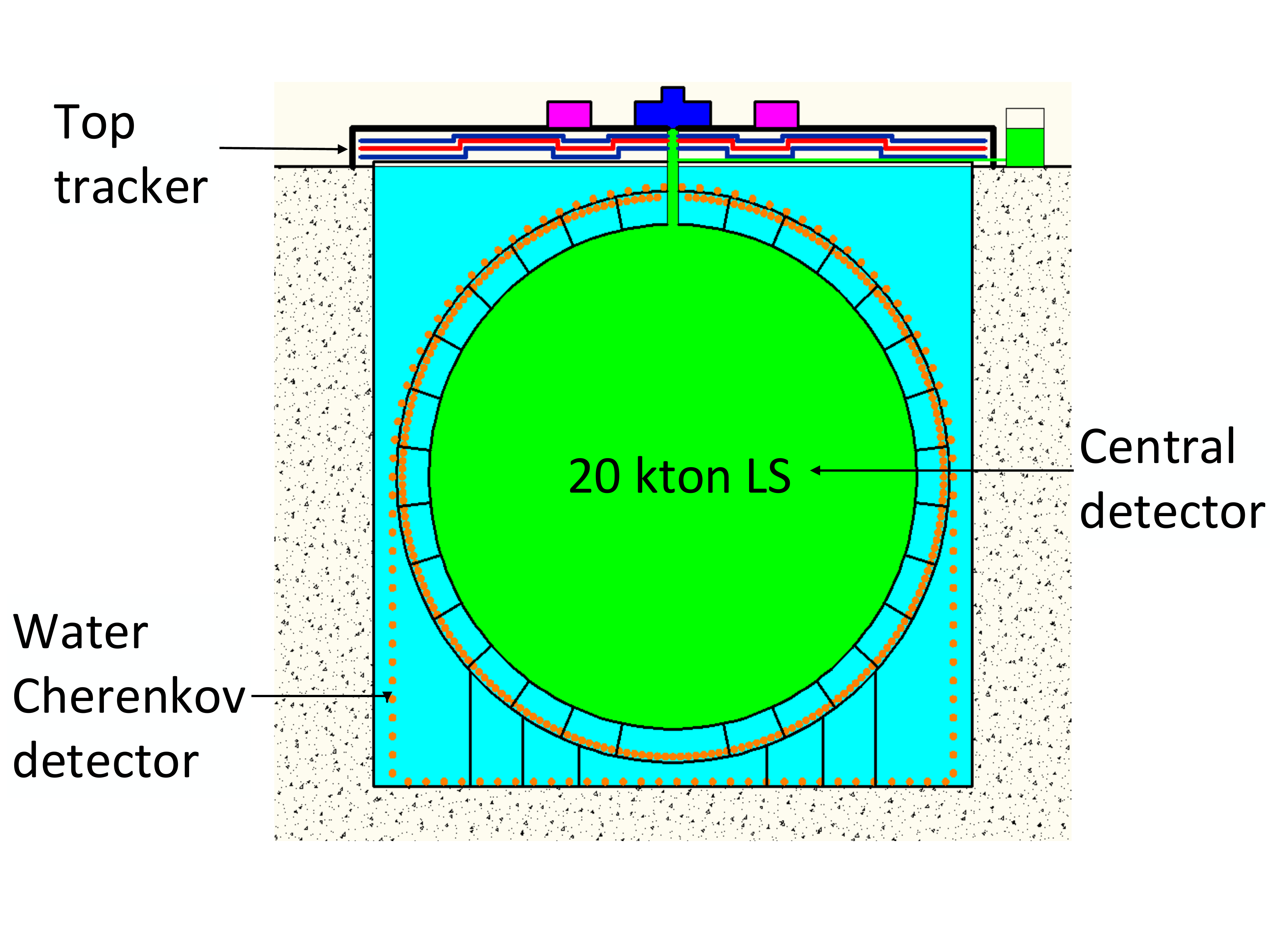}
\caption{Schematic view of the JUNO detector}
\label{fig1}
\end{figure}

The electronics and readout system need to handle the signals from the PMTs. The flash ADC (FADC) is proposed to be used to digitize the signal of fired PMTs with 1~GHz sampling rate. All the data from the FADC are transmitted to the online DAQ computing farm through data links.

The trigger system plays the role of rejecting backgrounds and keeping signal events with a sufficient high efficiency. A common trigger strategy is to select events using the multiplicity of fired PMTs, which has been successfully used by the Daya Bay experiment \cite{lab3}. It can reach 100\% trigger efficiency for inverse beta decay (IBD) events with 0.5 MeV trigger threshold, which can satisfy the requirements of neutrino mass hierarchy measurement. However, for some physics topics, like the study of solar neutrinos and supernova neutrinos, a much lower trigger threshold is preferred, which is severely limited by the coincidence of PMT dark noises and events rate of natural radioactivity. So, a more sophisticated trigger method is proposed based on vertex fitting, which will be discussed in detail in the next section.
                
\section{Trigger schemes of CD}

The common multiplicity trigger, used in Daya Bay experiment, is one of the trigger options in JUNO CD. Its typical trigger time window is 300 ns, determined by the decay time of LS \cite{lab4} and variations of time of flight (TOF) from events positions to each PMT. However, due to the relatively high dark noise rates (typically 50 kHz for MCP-PMT and 20 kHz for dynode PMTs) and large quantity of PMTs, a high trigger threshold has to be set in order to eliminate the coincidence of PMT dark noises. While the rate (R) of PMT dark noise coincidence can be accurately calculated using the formula (\ref{eq1}):

\begin{equation}
   R = \frac{1}{\tau}\sum_{i=m}^{N}iC^{i}_{N}(f\tau)^i
	(1-f\tau)^{N-i}
\label{eq1}
\end{equation}

where, $N$ is the number of PMTs in total and $m$ is the number of fired PMTs that determines the trigger threshold, assuming the common multiplicity trigger is used; $f$ denotes dark rate of each PMT; $\tau$ indicates length of trigger time window. For CD, a trigger time window of 300 ns is essential to keep the high trigger efficiency for physical events, and the calculated events rate caused by dark noise coincidence versus trigger threshold with 300 ns (dashed lines) and 80 ns (solid lines) trigger time window are shown in Fig. 2, in which the re-trigger is allowed and the dark noise rate of each PMT is assumed to be 50 kHz (blue lines) and 30 kHz (red lines) respectively. The PMT dark noise is one of the factors to determine the lower limit of the trigger threshold, and the shorter trigger time window can significantly reduce the events rate from PMT dark noise coincidence.
		
\begin{figure}[H]
\centering
\includegraphics[width=8cm]{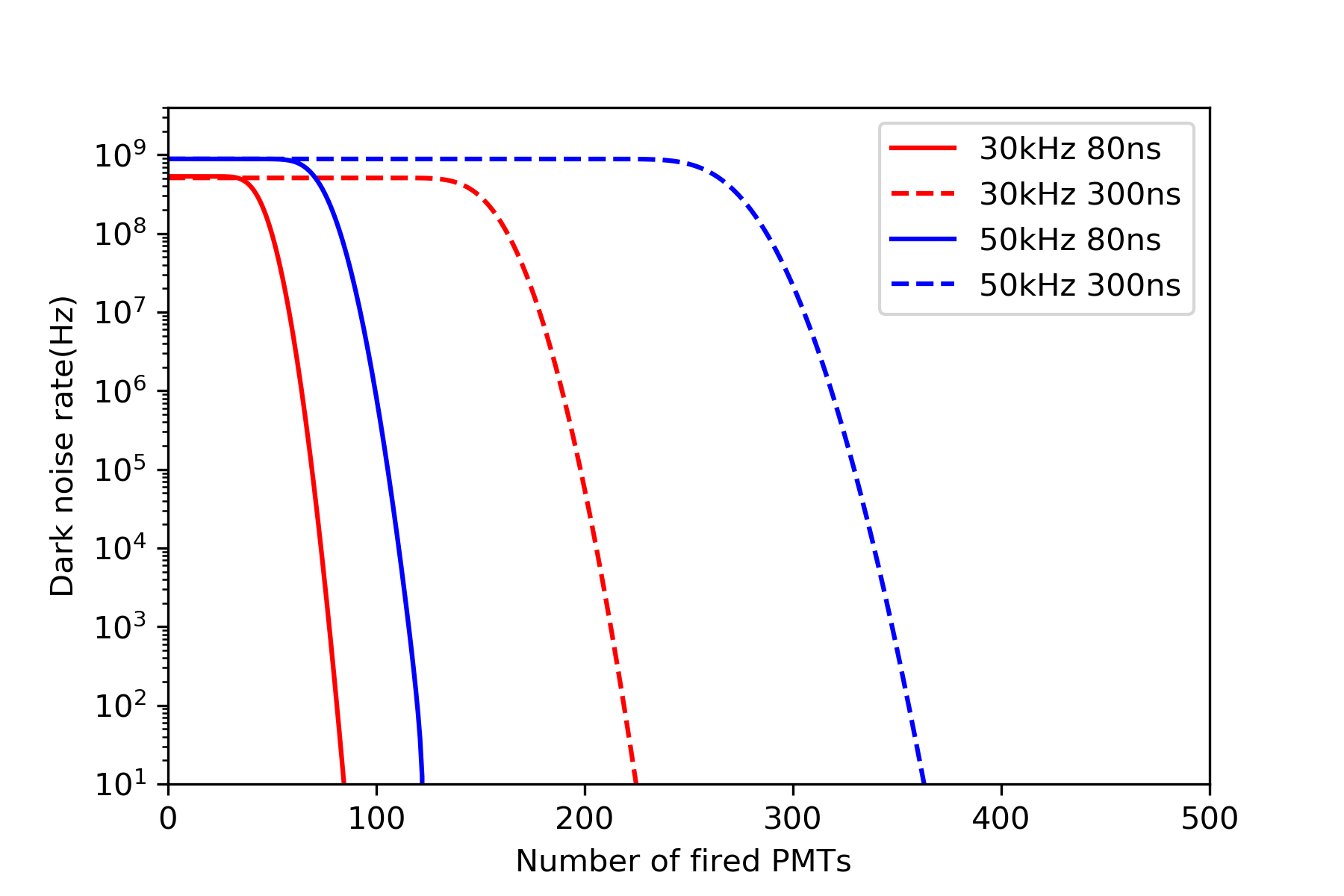}
\caption{Events rate caused by dark noise coincidence versus number of fired PMTs}
\label{fig2}
\end{figure}
		
In order to decrease the impacts from PMT dark noise, a sophisticated trigger scheme is proposed. The simple idea is to shorten the trigger time window to reduce the probability of PMT dark noise coincidence. One effective method to achieve this goal is to reconstruct events positions, then correct the TOF for each PMT with the known information of events positions and the PMT position. In TOF calculation, the equivalent velocity of light in liquid scintillator can be easily calculated by using $\frac{c}{n}$, where $c$ represents the speed of light in vacuum and $n$ denotes the refractive index of liquid scintillator at the peak emission wavelength of 430~nm. Since FPGA can not handle complex position reconstruction algorithms, a simple method is proposed instead to reconstruct events positions with an acceptable position resolution. In this method, CD is divided into 179 cubic volumes. For each volume, the dimension is about 5 m $\times$ 5 m $\times$ 5 m. To get the event position, the TOF is corrected for each PMT accordingly by looping through all volumes. If a volume contains the event, the narrowest first hit time distribution on fired PMTs should be gotten. Then, the center of this cubic volume is taken as the event position. And the radius position of this event (R) can be calculated. This method has been proved and validated by simulations. For demonstration, Fig. 3 shows the first hit time distribution on PMTs in different situations, by simulating 4 MeV e+ at (15m, 0, 0) in CD. After applying TOF correction in the volume which contain the generated position of event, a much narrower distribution is obtained (the blue line). Without TOF correction, the distribution is wider (the green line). For volumes which don't contain the generated position of event, hit time distribution is also broader (the red line). 
		
For the real situation, it is not necessary to perform the position reconstruction algorithm, to find out the correct volume which contains the event, because FPGA can run in parallel. For each volume, FPGA can make the trigger decision independently. Take one volume as an example, the TOF correction map for this volume is calculated and saved in register in advance. Since the trigger system runs along with a clock system, and the clock cycle  is 16~ns in JUNO, in every clock cycle, trigger system receives a collection of digital signals of fird PMTs from front-end electronics. Then, FPGA loads the TOF correction map of this volume and correct the hit time for each PMT. After that, a trigger decision time window of 80~ns slides along the time axis with a step of 16~ns, to check if the number of fired PMTs in the time window exceeds the pre-configured threshold. If it is true, a trigger signal will be sent out to the front-end electronics and data acquisition system, at the same time, the information of event position is also known (position of triggered volume). The above processes are performed for all volumes in parallel by FPGA. We also studied other trigger time windows (48ns, 64ns) and found the similar performance. 
		
With the new trigger scheme, a fiducial volume cut can be easily applied and it is very effective to reduce events rate of radioactivity, because most of radioactivity are from outer region of CD. The reconstructed positions determined by the new trigger scheme are also validated by using the truth positions gotten from Monte Carlo (MC) truth information. The bias between reconstructed and truth events positions is drawn in Fig. 4, by using $^{14}$C events uniformly distributed in LS with energy larger than 0.1 MeV. The largest bias in Fig. 4 is smaller than the volume size, 5 meters, which means the volume can be 100\% correctly found. The resolution of reconstructed positions is about 1.5 meters, which agrees with expectations.

\begin{figure}
\centering
\includegraphics[width=8cm]{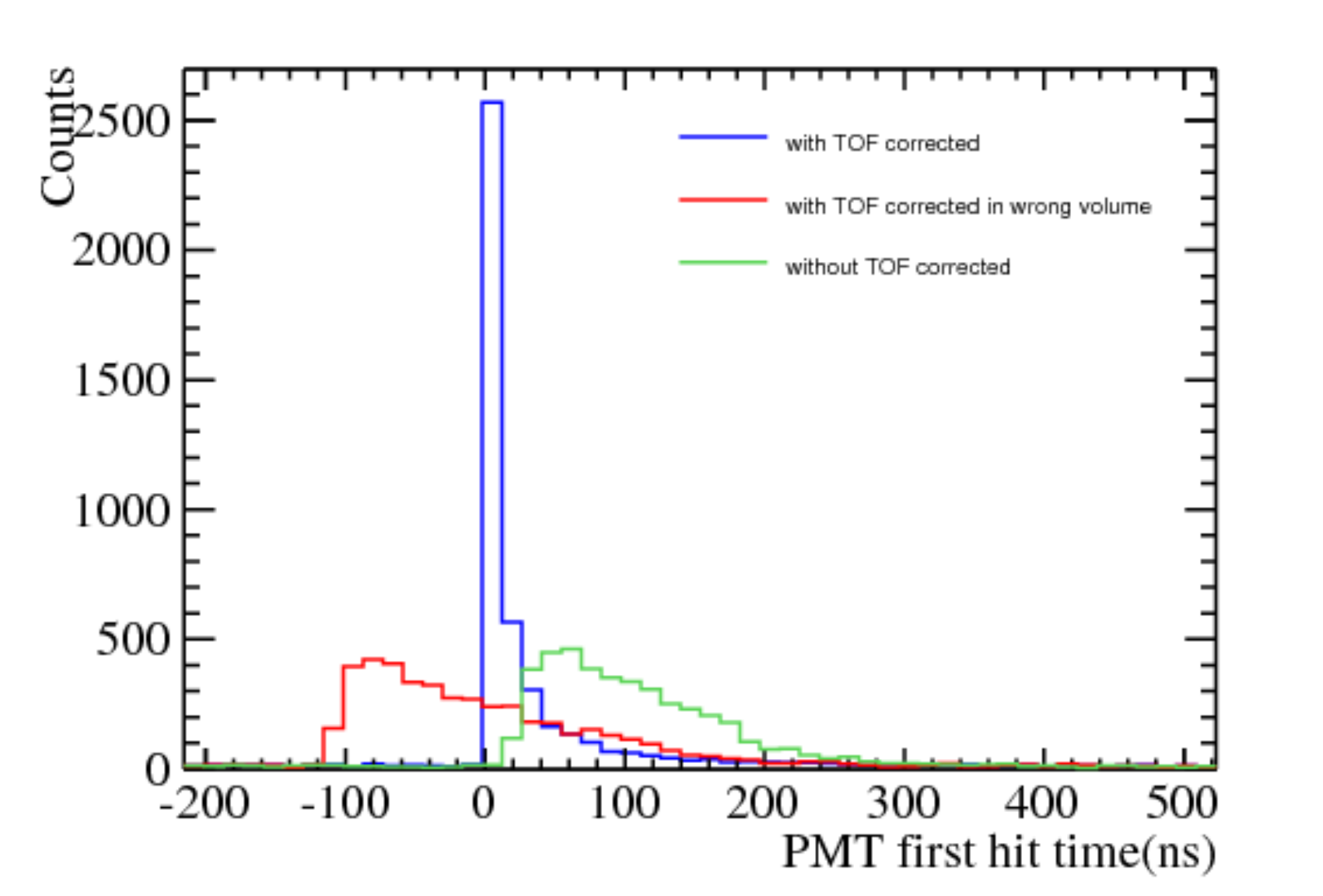}
\caption{First hit time of fired PMTs with different situations}
\label{fig3}
\end{figure}

\begin{figure}
\centering
\includegraphics[width=7cm]{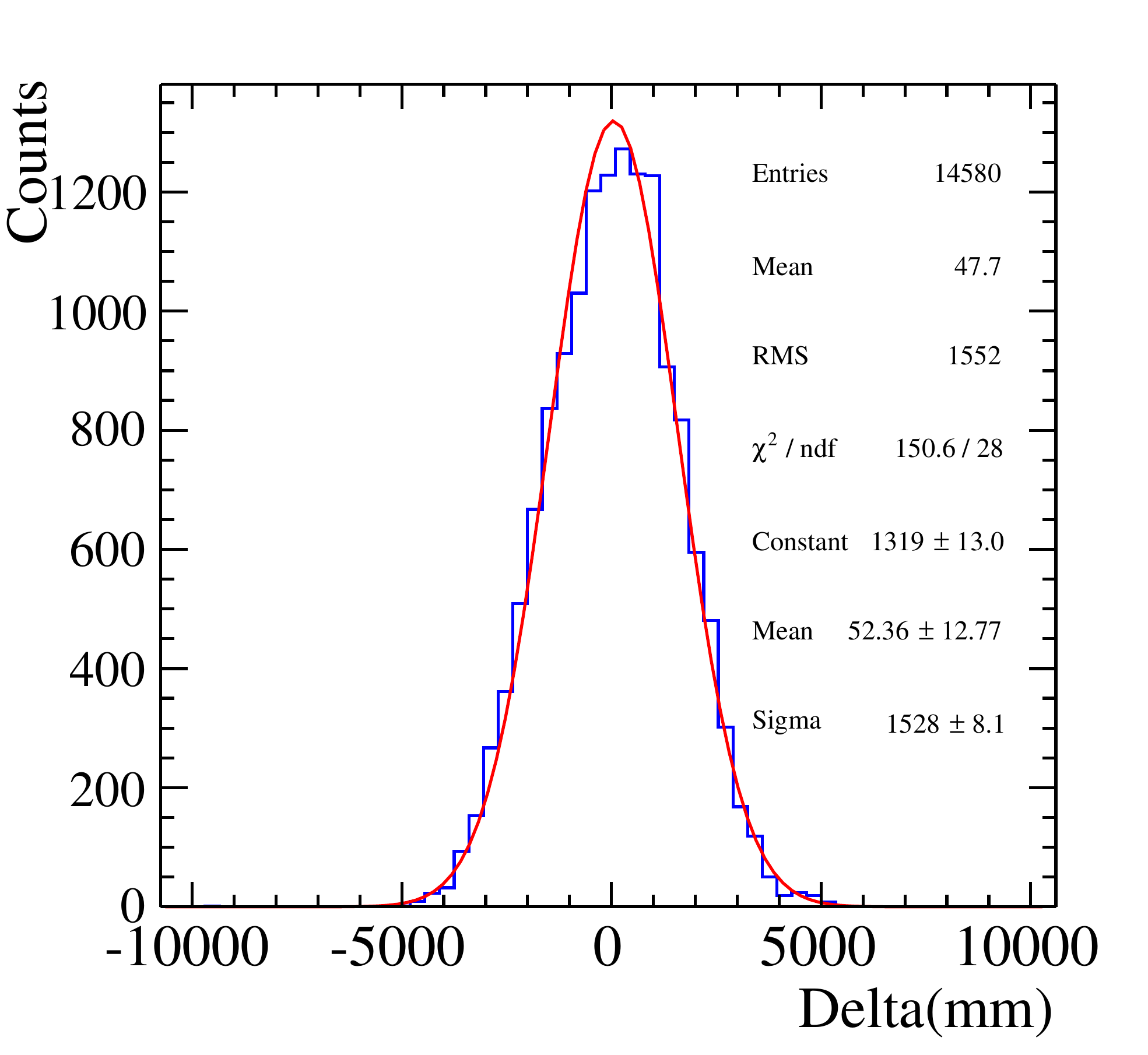}
\caption{Differences of true vertices and reconstructed vertices for $^{14}$C events}
\label{fig4}
\end{figure}

The performance of the new trigger scheme is carefully studied by using a detailed CD simulation software, which has been designed and implemented based on GEANT4 \cite{lab5} and SNiPER framework \cite{lab6}. Both the common multiplicity trigger method and the sophisticated trigger method have been implemented in the simulation software, and they are optional for users. By simulating gamma particles, uniformly distributed in CD, with three different energies (0.1 MeV, 0.2 MeV and 0.3 MeV) to mimic the physical signals, the numbers of fired PMTs with two different trigger schemes are shown in Fig. 5 (a) and Fig. 5 (b) respectively. In addition, the events rate of dark noise coincidence is also shown in same plots represented by dark lines. For both cases, the 50 kHz dark noise rate is assigned for each PMT. By comparing the two trigger schemes, the sophisticated trigger shows much better performance, with this trigger method, the effect from PMT dark noise is eliminated even for 0.1 MeV physical events, but this is not the case for the common multiplicity trigger strategy.
        
\begin{figure}
\centering
\subfigure[]{\includegraphics[width=7cm]{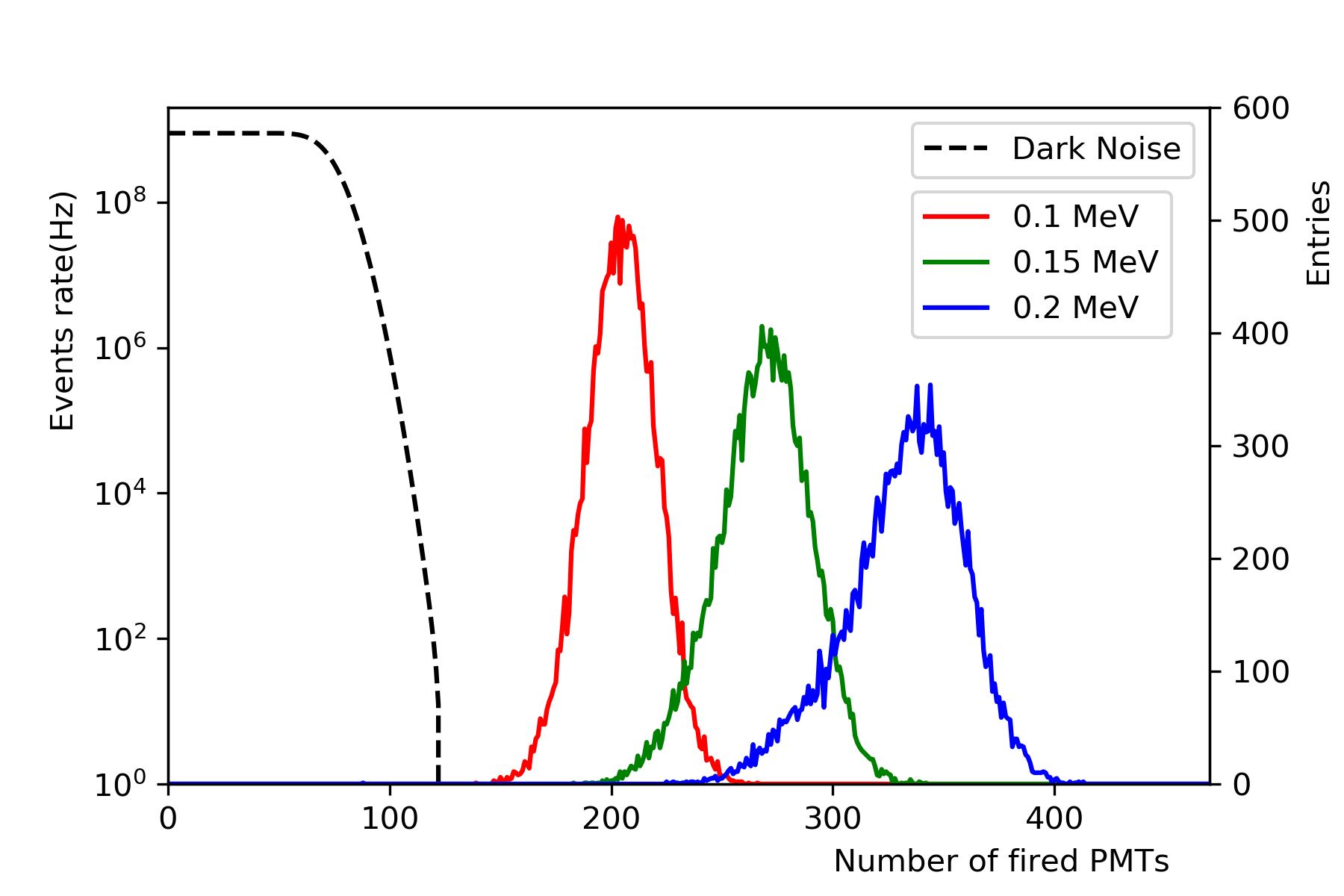}}
\subfigure[]{\includegraphics[width=7cm]{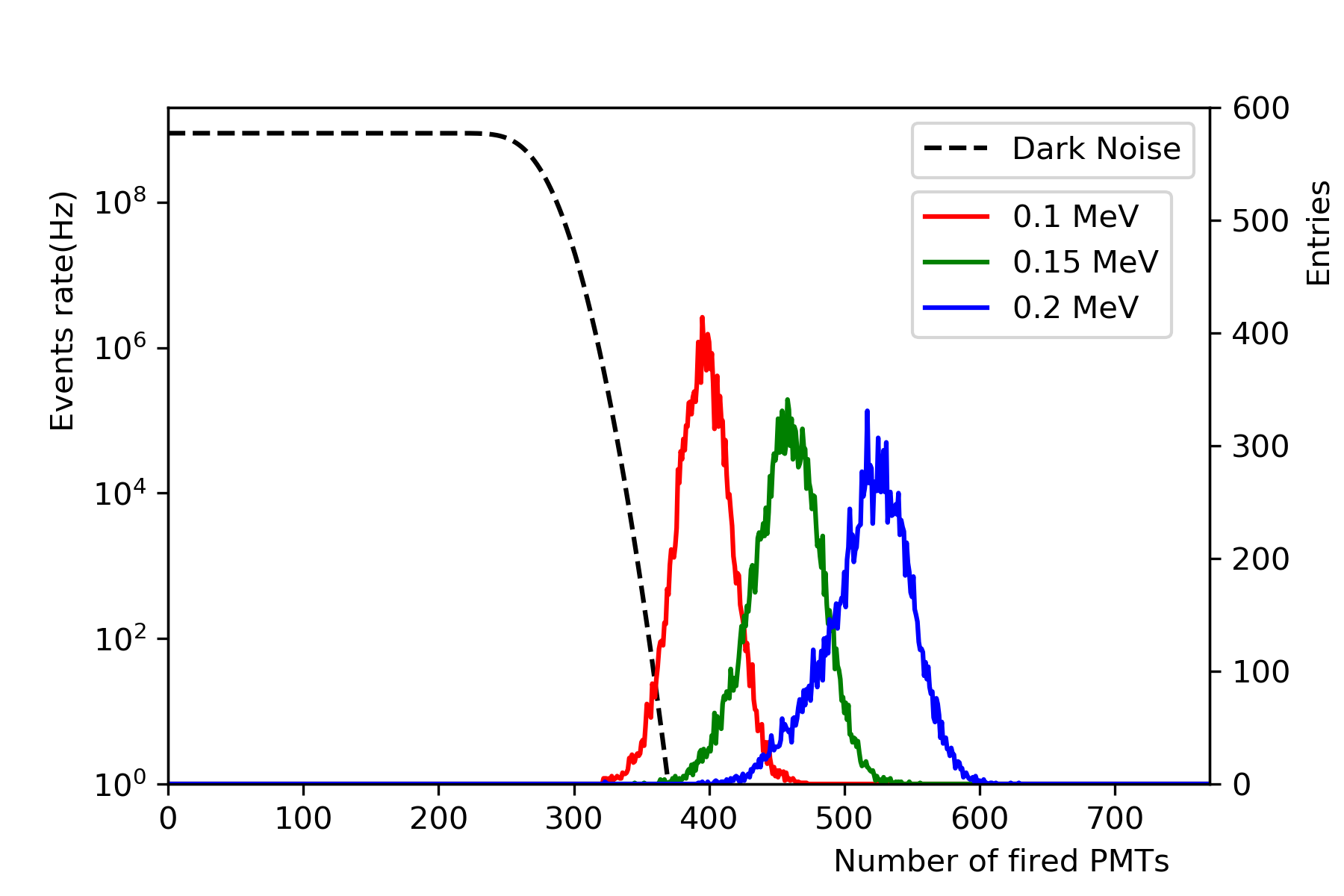}}
\caption{Maximum number of fired PMTs in trigger time window of 80 ns (a) and 300 ns (b) for gamma particles with different energies and events rate of coincidence of dark noise.}
\label{fig.5}
\end{figure}
        
In JUNO, the lower trigger threshold is preferred, but it is limited by the events rate during data taking. As known, events rate from radioactivity will increases when trigger threshold decreases. So it is important to study the events rate from different sources and check its impacts on trigger threshold. This work is done based on full detector simulation with two different trigger schemes.

\section{Events rate for physical events}
The events detected by CD can be divided into physical events and background events. Physical events include all events of concern, like reactor neutrino, solar neutrino, geo-neutrino, atmospheric neutrino, etc.
For each physical channel, the original events rate has been well estimated in the JUNO yellow book \cite{lab7}. Events rate of different event types without trigger threshold are summarized in Table 1. The solar neutrino accounts for the major part of physical events, which can reach 0.45 Hz totally if no trigger threshold is applied, mainly contributed from pp channel, where the energy of pp neutrino is quite low.

\begin{table}
\centering
\caption{ \label{tab1}  Events Rate of different event sources} 
%\footnotesize
\smallskip
\begin{tabular*}{80mm}{@{\extracolsep{\fill}} c c }
\hline
  Event source&Rate  \\
\hline
				IBD event               & 83 per day   \\
				Li9/He8 $\beta$-n decay & 84 per day   \\
				fast neutron            & 0.1 per day  \\
				C13($\alpha$,n)O16      & 0.05 per day  \\
				Geo-neutrino            & 1.5 per day  \\
				cosmic muon             & 3 Hz          \\
				Natural Radioactivity   & $\sim$1 MHz  \\
				Solar neutrino          & 0.45 Hz       \\
				Atmospheric neutrino    & 0.94 per day  \\
				\hline
				%\bottomrule
\end{tabular*}
\end{table}

The total events rate from physical channels is less than 1 Hz, which is negligible comparing with those from radioactivity backgrounds. However, if a supernova bursts at a typical distance of 10 kpc, neutrino events rate has been estimated to be hundreds of Hz from different channels, and events rate is much higher in the first second during the explosion. The supernova can last for about 10 seconds. For a supernova with shorter distance, the higher events rate is expected. The trigger scheme for supernova neutrino is under development in JUNO. 

\section{Events rate for background events}

\subsection{Cosmic muons}
JUNO detector is located underground with about 700 m overburden. By considering the detailed mountain map at JUNO site, the cosmic ray flux can be simulated by MUSIC \cite{lab8}. Then, the muon flux in experimental hall is obtained to be 0.003 Hz/m$^2$, and the average energy of muons are determined to be about 215 GeV. Based on the muon flux and dimensions of CD, the total muon events rate in CD is calculated to be 3 Hz.

The cosmic muons which go into the CD can be separated into shower muons and non-shower muons. The non-shower muons are minimal ionization particles. The shower muons may generate neutrons and radioactivity isotopes with long lifetime, which can form delayed signals and be tagged as separated events besides muon track itself. A detailed muon simulation shows that about 20\% of muons are shower muons and the events rate induced by muons is about 5 Hz totally.

\subsection{Radioactivity background}
One of the main contributions to events rate is the natural radioactivity background, which comes from various sources, like $^{238}$U, $^{232}$Th ,$^{40}$K. 
They come from different materials in JUNO detector. In order to control the events rate of background, the purity of each material used to make detector has to be screened and the requirements on radio-purity are driven by the sensitivity of neutrino mass ordering. In Daya Bay experiment, the U/Th concentrations in LS are 2$\times10^{-14}$ g/g and 4$\times10^{-14}$ g/g respectively. In JUNO, a better purification system will be applied, which can reduce the U/Th by two orders of magnitude, to the level of $10^{-16}$ g/g. For $^{40}$K, it can reach $10^{-17}$ g/g. 
Besides U/Th/K, cosmic muons continuously produce $^{14}$C from $^{14}$N, via the reaction $^{14}$N(n,p)$^{14}$C.
If the petroleum used to make LS is located deep underground and is shielded from high cosmic ray flux, one can expect very low $^{14}$C contaimination in LS. 
But for the worst case, petroleum may locate on the surface of earth which suffers very high flux of cosmic muons, $^{14}$C abundance becomes higher. In this work, we set $^{14}$C abundance to be $10^{-18}$ g/g. The events rate induced by $^{14}$C can be easily scaled for other concentrations. The radioactivity in PMT glass depends on which glasses are selected to make PMTs. In JUNO, U/Th/K concentrations in MCP-PMT glass are 2.5 Bq/kg, 0.5 Bq/kg and 0.5 Bq/kg respectively, for Hamamatsu PMTs, the radio-purity is several times worse than that of MCP-PMTs \cite{lab9}. However, due to 1.4-meter water buffer between PMTs and LS, the events rate from PMT glass can be reduced a lot. This effect has been included in our simulation. The acrylic sphere is quite closed to LS in which the radioactivity can easily deposit their energies in LS and be triggered to form events. Radioactivity existing in stainless truss also contributes total events rate in CD, but similar with that in PMT glass, it is heavily suppressed by water buffer.
The upper limits of U/Th/K concentrations in stainless steel truss are 12.4 mBq/kg, 20 mBq/kg and 54.2 mBq/kg respectively based on reference \cite{lab10}. $^{60}$Co can also be generated when cosmic muons go through stainless truss. Radon in water is required to be less than 0.2 Bq/m$^3$, in this way, there is no impact on the mass hierarchy sensitivity. And U/Th/K concentrations in rock are 10ppm, 30ppm and 5ppm respectively. The requirements of upper limits to radioactive impurities in different detector materials are summarized in Table 2. 

	\begin{table}
	\centering
		\caption{ \label{tab2}  The requirements of upper limits to radioactive impurities of different detector materials.}
		\smallskip
		\footnotesize
		\resizebox{\textwidth}{!}{
		\begin{tabular*}{160 mm}{@{\extracolsep{\fill}}ccccccccc}
		%\begin{tabular}[40 mm]{|c|c|c|c|c|c|c|c|c|}
		\hline
			\hphantom{0} & $^{238}$U & $^{232}$Th & $^{40}$K  & $^{210}$Pb & $^{85}$Kr & $^{39}$Ar & $^{60}$Co & $^{14}$C\\
			\hline
			LS & $10^{-6}$ ppb & $10^{-6}$ ppb & $10^{-7}$ ppb & $1.4\cdot10^{-13}$ ppb & 50 $\mu$Bq/m$^3$ & 50 $\mu $Bq/m$^3$ & \~{}  & 1$\times10^{-18}$g/g \\
			Glass & $257$ ppb & $200$ ppb & $13.8$ ppb & \~{} & \~{} & \~{} & \~{} & \~{} \\
			Acrylic & $4.9$ ppt  & $81$ ppt & $3$ ppt & \~{} & \~{} & \~{} & \~{} & \~{} \\
			Steel & $ 12.4$ mBq/kg & $20$ mBq/kg  & $54.2$ mBq/kg & \~{} & \~{} & \~{}& $2 $ mBq/kg  & \~{} \\
			\hline
			%\bottomrule
		\end{tabular*}}
	\end{table}

	%\begin{multicols}{2}

\subsection{Results of radioactivity background simulation}

For each material in JUNO detector, the radioactivity listed in Table 2 are used as inputs in detector simulation to study its events rate. The results with different trigger thresholds are shown in Fig. 6 (a) and Fig. 6 (b), corresponding to results of the sophisticated trigger scheme and the common multiplicity trigger, respectively. The trigger efficiencies of physical events are indicated by the solid lines, determined by simulating gamma particles with three different energies and uniformly distributed in CD. The dark noise of PMTs has been taken into account in the simulation and assumed to be 50 kHz for each PMT.
		
\begin{figure}
\centering
\subfigure[]{\includegraphics[width=7cm]{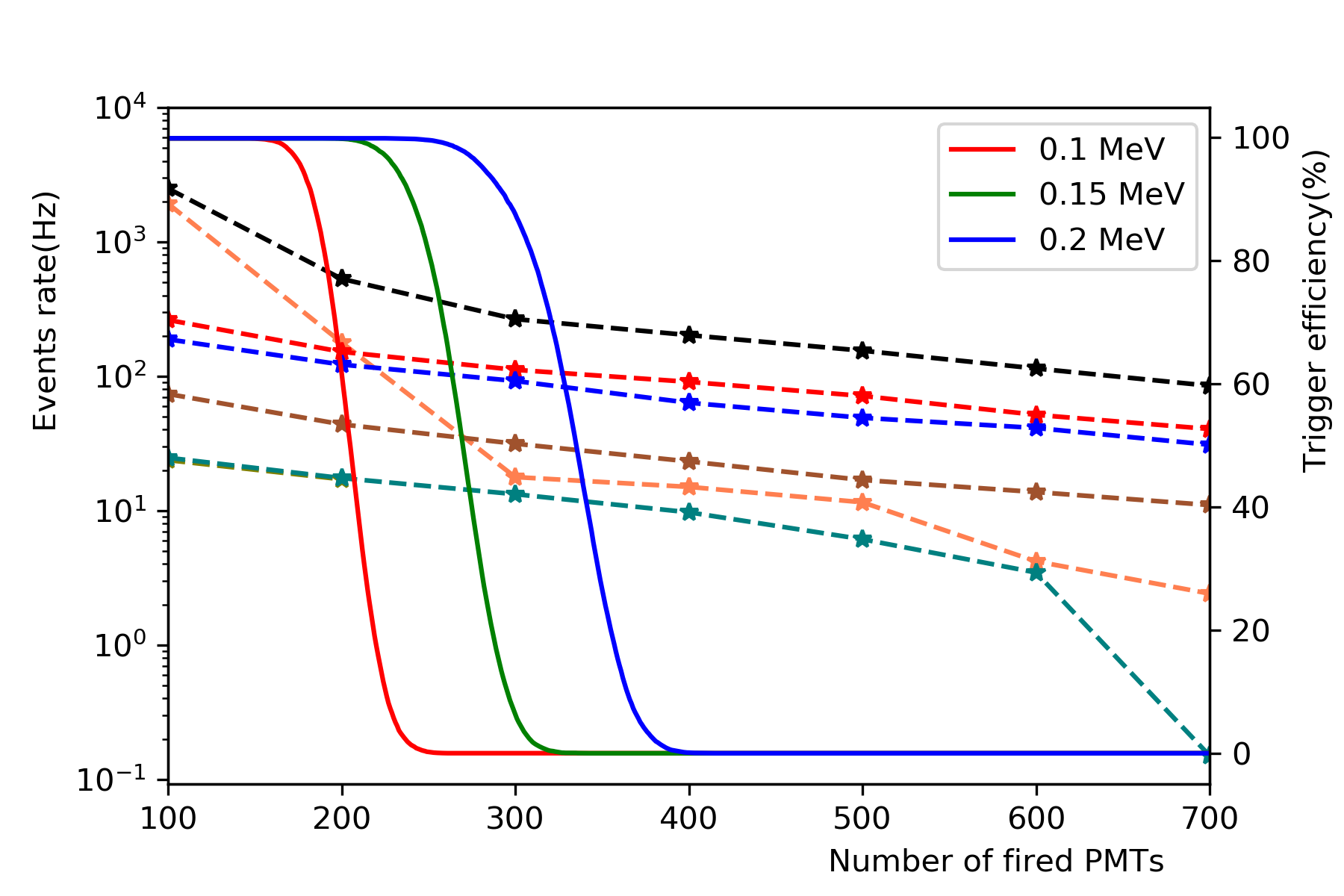}}
\subfigure[]{\includegraphics[width=7cm]{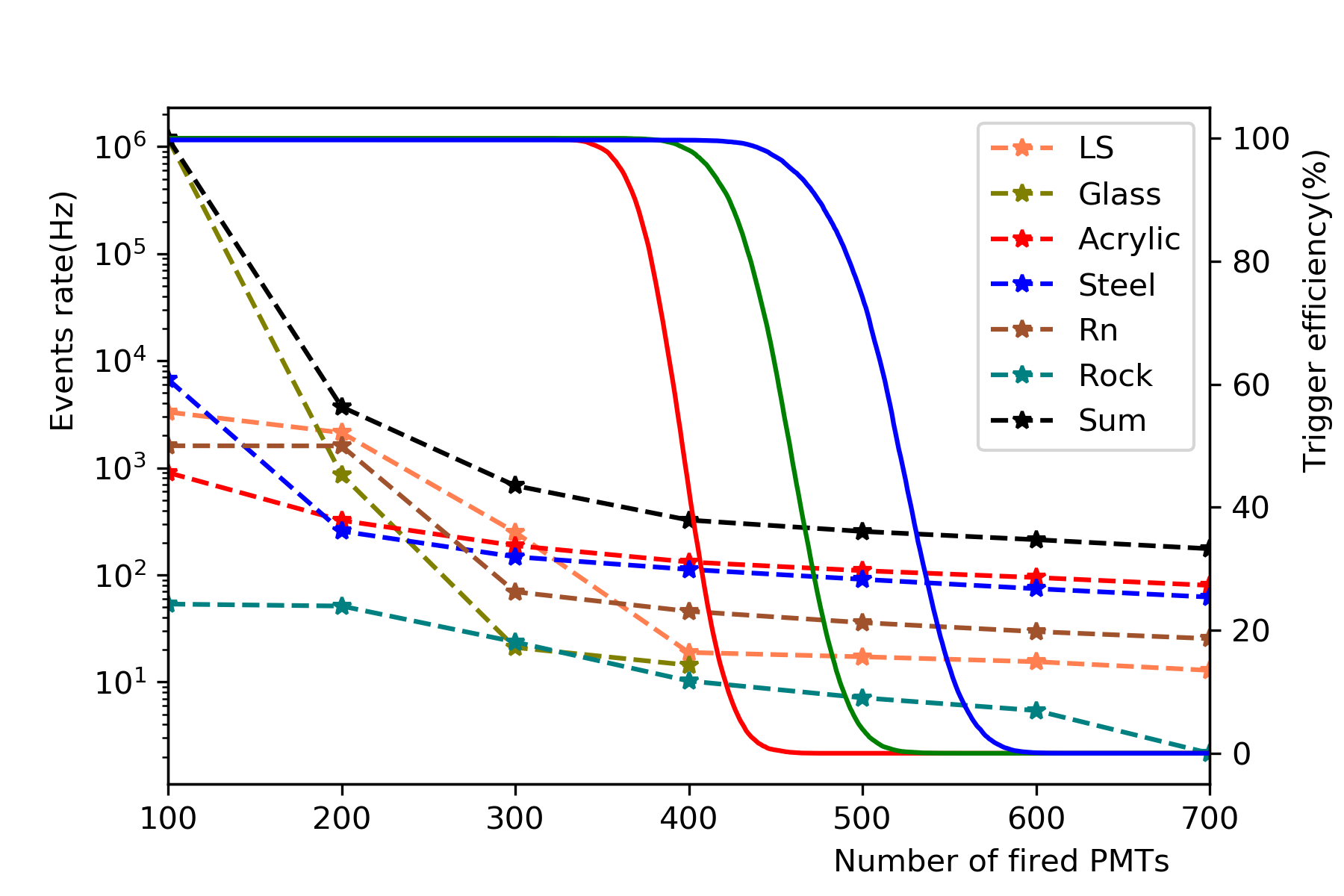}}
\caption{Events rate contributed by different materials in CD and trigger efficiency of physical events with different energies. (a) is from sophisticated trigger and (b) for common multiplicity trigger}
\label{fig.6}
\end{figure}

The events rate increases when the trigger threshold decreases. Assuming DAQ can accept 1 kHz events rate during data taking, then based on Fig. 6, the trigger threshold can be set at 150 and 380 for sophisticated trigger and common multiplicity trigger respectively. In this case, sophisticated trigger can keep 100\% trigger efficiency even for 0.1 MeV physical events, but the efficiency of common multiplicity trigger is ~60\% for 0.1 MeV events. The trigger threshold is limited by the radio-purity of LS, which is dominated at low energy region and $^{14}$C is the main contributor. $^{14}$C goes through radioactive beta decay by emitting an electron and an electron anti-neutrino. The emitted electrons have a maximum energy of 0.156 MeV and only a small fraction ($<$ 0.2\%) of $^{14}$C can be triggered with 0.2 MeV threshold due to energy smearing and pileup. While, a fraction of $^{14}$C can be triggered by 0.1 MeV threshold. At high energy region ($>$ 0.25 MeV), the acrylic and radon are two major sources of events.
		
Since most of the radioactivity background comes from the outer region of CD, and the sophisticated trigger scheme has the ability to select events in different fiducial volumes, the events rate coming from backgrounds can be further reduced. The events rate from different radioactivity sources are summed up and the results are listed in Table 3 with different trigger threshold and in different fiducial volumes. The trigger threshold is defined as number of fired PMTs and the fiducial volumes are selected by applying different radius cuts. Based on energy scale in CD ($\sim$1200 p.e./MeV), the threshold can be converted to visible energy, however in this process, the coincidence of PMT dark noise should be subtracted before the conversion. In Table 3, the energy threshold of 200 corresponds to about 0.1 MeV visible energy where 50 kHz dark noise rate for each PMT is assumed. The events rate decreases when the fiducial volume becomes smaller, since part of radioactivity comes from the outer region of CD, on the other hand, both the radioactivity in LS and events rate of physical events are reduced due to the smaller target mass. This is a trade-off between trigger threshold and statistics of signals.

	\begin{table}
	\centering
		\caption{ \label{tab3} The sum radioactivity single rate in central detector with different trigger threshold and fiducial volumes}
		\footnotesize
		\smallskip
		\begin{tabular*}{150mm}{@{\extracolsep{\fill}} c c c c c c c}
		\hline
			 Singles Rate (Hz)&Thres.$>$100 & Thres.$>$200 & Thres.$>$300 & Thres.$>$400 & Thres.$>$500 & Thres.$>$600 \\
			\hline
			
			R$<$10m  &1292  &126  &15   &11   &9    &6  \\
            R$<$11m  &1338  &134  &19   &14   &11   &8  \\
            R$<$12m  &1483  &167  &35   &26   &21   &15 \\
            R$<$13m  &1614  &200  &54   &40   &31   &23 \\
            R$<$14m  &1614  &200  &54   &40   &31   &23 \\
            R$<$15m  &1703  &228  &71   &53   &42   &31 \\
            R$<$16m  &2104  &361  &152  &116  &90   &66 \\
            R$<$17m  &2320  &439  &202  &154  &120  &88 \\

             \hline
			%\bottomrule
		\end{tabular*}
	\end{table}

\section{Conclusion}
To detect low energy events is essential for JUNO to study the solar neutrinos and supernova neutrinos. With the proposed sophisticated trigger scheme, the effect of dark noise coincidence is significantly suppressed. The trigger energy threshold can be reduced to 0.2 MeV, and if the concentration of $^{14}$C in LS can be well controlled, such as less than 10$^{-18}$ g/g, the trigger threshold can even reach to 0.1 MeV. The total events rate from physical channels excepting supernova events in JUNO is less than 1 Hz, which is negligible comparing with those from radioactivity backgrounds. The proposed trigger scheme also has the ability to trigger events in a certain fiducial volume, which can further suppress the radioactivity events coming from the outer region of CD. The results of the present study provide important events rate estimation and trigger scheme both for JUNO design and disk storage preparation. 
	
\acknowledgments
We gratefully acknowledge support from NSFC grant No.11875279 in China.

\end{document}